\documentstyle[graphicx,psfrag,amssymb,prl,twocolumn,aps]{revtex}

\newcommand{\bm}[1]{\mbox{\boldmath$#1$}}

\begin{document}
\draft
\twocolumn[\hsize\textwidth\columnwidth\hsize\csname @twocolumnfalse\endcsname
\title{Short wavelength spectrum and Hamiltonian stability of vortex rings}
\author{Uwe R. Fischer and Nils Schopohl}
\address{
Eberhard-Karls-Universit\"at T\"ubingen, 
Institut f\"ur Theoretische Physik \\
Auf der Morgenstelle 14, D-72076 T\"ubingen, Germany}
\date{\today}
\maketitle


\begin{abstract}                
We compare dynamical and energetical stability criteria for vortex rings.  
It is argued that 
vortex rings 
will be intrinsically 
unstable against perturbations with short wavelengths
below a critical wavelength,  
because the canonical vortex Hamiltonian is unbounded from below for these
modes. 
To explicitly demonstrate this behaviour, we derive the 
oscillation spectrum of vortex rings in incompressible, inviscid  
fluids, within a 
geometrical cutoff procedure for the core. 
The spectrum develops  
an anomalous branch of negative group velocity, 
and approaches the zero of energy for  
wavelengths which are 
about six times the core diameter. 
We show 
the consequences of 
this dispersion relation 
for the 
thermodynamics of vortex rings in superfluid $^4\!$He at low temperatures.  
\end{abstract}

\


\

] \narrowtext
\section{Introduction}
The long wavelength 
oscillation 
spectrum of large vortex rings in incompressible, inviscid fluids 
is established since
the pioneering work of W. Thomson (Lord Kelvin) \cite{Kelvin},  
J. J. Thomson \cite{JJThomson}, and Pocklington \cite{pocklington}.
The validity of that spectrum 
is restricted to wavenumbers much less than the inverse core size and 
rings which are large compared to the extension of the core.
There are, however, processes for which it is desirable to know the 
large wavenumber 
properties of the spectrum for smaller rings: 
Vortex ring nucleation, reconnection  of vortex filaments 
and dissipation in the turbulent energy cascade 
are believed to occur on very small length scales, reaching down to 
a few times the vortex core size.
Because a line vortex represents a string object, elasticity modes 
will be excited during the rapid movements executed by the string on small 
length scales. 
A fluctuating line should have 
equilibrium states different from a non-fluctuating one, because the quantum 
or classical statistical fluctuations renormalize the total 
free energy as compared to the undeformed ring.

In what follows, 
we shall derive the collective, small amplitude 
oscillation modes of a vortex ring in an incompressible, inviscid fluid. 
The dispersion relation 
is exact within the geometrical cutoff procedure we employ,  
and displays a maximum and an anomalous branch of negative group velocity.
The critical wavelength for the spectrum 
to possess a positive excitation energy 
corresponds to one oscillation of line 
within a length about an order of magnitude above 
a geometrically defined core size. 
It will be argued that, due to the properties of this spectrum and 
the structure of the canonical Hamiltonian, a vortex ring is 
potentially unstable in an intrinsic manner, because the Hamiltonian 
is unbounded from below for short wavelengths. 
Physically, the energetical instability is caused  
by the fact that quantities playing the role of ``mass''  and 
``spring constant'' in the Hamiltonian do simultaneously
assume negative values. 
The energetical instability, taking place
for large perturbations which are of wavelengths less than 
the critical wavelength 
(of about six times the core diameter in our core model), 
occurs though the ring is dynamically stable for nearly 
all wavelengths down to the core size. 
The dynamical instability of a vortex ring, related to the ocurrence of 
imaginary excitation frequencies, is only 
relevant for certain critical wavelengths. We will thus present in this paper
an argument that the relevant stability criterion 
for a vortex should be that of energetical stability.

Below, in the section to follow, 
we will first introduce an action principle which gives a 
transparent representation of the vortex dynamical behaviour in 
incompressible, inviscid fluids, from which the canonical 
Hamiltonian for 
small perturbations in section \ref{Hamiltonian}, 
representing the vortex eigenmodes, naturally follows.
Section \ref{Thermodynamics} gives an account of the thermodynamics
of vortex rings, related to the excitation spectrum on a quantized vortex 
in superfluid $^4\!$He, where the consequences of the predictions made
in this work should be particularly clearly seen. We conclude with 
some remarks.

 
\section{Derivation of the oscillation modes}
\subsection{Action Principle} 
The peculiarity of the 
dynamical behaviour of vortices in the incompressibility approximation,   
the fluid having a constant mass density $\rho_0$, 
consists in the fact that configuration space and phase
space coincide \cite{poincare,onsager}. The momenta are, in this limit, 
generally expressible as functions of the co-ordinates, and play no 
independent dynamical r\^ole.  
This fact gives rise to the following action functional of the line 
configuration $\cal C$, 
in terms of the positions ${\bm R}$ of line elements $d{\bm R}$    
with a constant velocity circulation $\Gamma$ \cite{Regge}:
\begin{equation}
S\left[ {\cal C}\right] =\int_{0}^{t}\!dt\left( -\frac{\Gamma }{3}\,\rho_0
\,\oint_{{\cal C}}\,\left\langle d{\bm R\wedge {\bm R}
\;,\;\partial }_{t}{\bm R%
}\right\rangle -H\left[ {\cal C}\right] \right). \label{vortex action}
\end{equation}
The factor $\frac{1}{3}$ in the kinematical term reflects our choice of
cartesian co-ordinates in what follows, and corresponds to a (co-ordinate) 
gauge for the vortex momentum \cite{annalspaper}. 
The vortex kinetic energy is given by the Biot-Savart expression  
\begin{equation}\label{BS energy}
H\left[ {\cal C}\right] =\frac{\Gamma ^{2}}{2}\,\rho_0 \,\oint_{{\cal C}%
}\oint_{{\cal C}}\frac{1}{4\pi }\frac{\left\langle d{\bm R},d{\bm R}^{\prime
}\right\rangle }{\left| {\bm R}-{\bm R}^{\prime }\right| }\,,
\end{equation}
where the shorthand notation ${\bm R}={\bm R}\left( \phi ,t\right) $ 
and ${\bm R}^{\prime }={\bm R}\left(
\phi ^{\prime },t\right) $ is used.
This relation for the energy 
yields the usual asymptotic logarithmic dependence of the stationary
vortex energy on the infrared cutoff $L$ (system size, distance to the next 
vortex, or 
radius of a vortex ring) 
and the ultraviolet cutoff $\xi_c$ (the core size), in the form 
$\ln[L/(C\xi_c)]$, where $C$ is a core model dependent constant.  
Stationarity of the action for first order variation 
of the action (\ref{vortex action}) after ${\bm R}$ 
leads to the local velocity of a line element  
being perpendicular to the line element, and 
given by the Biot-Savart nonlocal induction law:
\begin{eqnarray}
\label{BS equation of motion} 
d{\bm R}\wedge\! \left( \partial _{t}{\bm R}-\frac{\Gamma }{4\pi }\oint_{{\cal %
C}}d{\bm R}^{\prime }\wedge \frac{{\bm {R-R}}^{\prime }}{\left| {\bm R}-{\bm R}%
^{\prime }\right| ^{3}}\right) ={\bm 0}\, .  \nonumber
\end{eqnarray}
The above proves that the action (\ref{vortex action}) leads to the correct
equations of motion, familiar from the fluid dynamics literature 
\cite{saffman}.  

Let ${\bm R}{\,}(\phi ,t)$, for $0\leq \phi \leq 2\pi $, describe the
instantaneous shape of a moving vortex ring at time $t$, fluctuating with
amplitude ${\bm u}(\phi ,t)$ around its circular {equilibrium} shape 
${\bm R}_{0}{\,}(\phi ,t)$, so that 
${\bm R}{\,}(\phi ,t) = {\bm R}_{0}{\,}(\phi ,t) +{\bm u}(\phi ,t)$. 
Then, we parametrize line element position, 
equilibrium position, and small 
perturbations around this equilibrium  as follows 
\begin{eqnarray}
{\bm R}\left( \phi ,t\right) &=&
{R}_{\,\perp}(\phi ,t)
\left( \cos \phi \,\widehat{{\bm e}}%
_{x}+\sin \phi \,\widehat{{\bm e}}_{y}\right)
+R_\parallel (\phi ,t)\,\widehat{{\bm e}}_{z} 
\nonumber \\
{R}_{\,\perp}& = &r_{0} + u_{\,\perp }\left(\phi ,t\right)\,,\quad  
{R}_{\,\parallel}= v_0 t +  u_{\,\parallel }\left(\phi ,t\right)\,.
\label{shape ansatz}
\end{eqnarray}
From this choice of co-ordinates 
and the form of the 
action (\ref{vortex action}), the phase space variables 
for small oscillations $\bm u(\phi,t)$ are concluded to be 
\begin{eqnarray}
q\left( \phi ,t\right) 
& = & u_{\,\parallel }\left(\phi ,t\right)
\nonumber \\
p\left( \phi ,t\right)  &=&(\Gamma \rho_0 r_0) u_\perp\left(\phi ,t\right)\,.
\label{canonical variables}
\end{eqnarray}
These phase space variables are employed for the description
of the vortex eigenmodes which follows. 
\subsection{The spectrum} 
The fundamental cutoff to be introduced 
for the continuum description to be valid is that the separation 
of two line elements should always exceed a length $\xi_c$,  
\begin{equation}\label{cutoffprescr} 
\left| {\bm R}-{\bm R}^{\prime }\right| >\xi _{c}\,.
\end{equation}
The length $\xi_c $ is thus defined as the {\em cutoff diameter}
 of the vortex core.  
The above prescription is the simplest exact procedure to ensure that 
the Biot-Savart integrals remain regular. 
If the Biot-Savart description is refined by, {\it e.g.},
a density profile in the core smoothly increasing within a distance 
$\xi_c/2$
to the constant $\rho_0$, instead of being cut off to be exactly 
zero at $\xi_c/2$, this will effectively yield a different ultraviolet cutoff, 
that is, a different core constant $C$ of order unity, 
multiplying a (fixed) value of $\xi_c$. However, the dynamical behaviour of the
vortex line on a scale well 
outside the core domain will not be affected by the core model.

To leading order in the fluctuations, the 
condition (\ref{cutoffprescr}) is equivalent to 
\begin{equation}
\left| \sin \frac{\phi -\phi ^{\prime }}{2}\right| >\frac{\xi _{c}}{
2r_{0}+u_{\bot }+ u^\prime_{\bot }} \,. 
\label{effective cut-off}
\end{equation}
Introducing the above condition in the integrals determining the velocity 
in (\ref{BS equation of motion}) by means of a Heaviside step function, 
we obtain the equilibrium velocity of the ring 
\begin{eqnarray}
v_{0} 
&=&\frac{\Gamma }{4\pi }\frac{1}{4r_{0}}\int_{\delta }^{2\pi -\delta }d\phi
^{^{\prime \prime }}\frac{{\small 1}}{\sin \frac{{\small \phi }^{\prime
\prime }}{{\small 2}}}  \nonumber \\
&=&\frac{\Gamma }{4\pi r_{0}}\ln \left( \frac{2r_{0}+2\sqrt{%
r_{0}^{2}-\xi _{c}^{2}}}{\xi _{c}}\right)\nonumber\\
&=&\frac{\Gamma }{4\pi r_{0}}\ln \left[
\cot(\delta/4)\right]\,,\label{v0}
\end{eqnarray}
where the cutoff angle is determined by the parameter 
\begin{equation}
\delta =2\arcsin \frac{\xi _{c}}{2r_{0}}\,.\label{defdelta}
\end{equation}
To obtain the ring oscillation modes, 
we expand the small quantities $u_\parallel$ and $u_\perp$
in a Fourier series, 
$u_{\bot }\left( \phi ,t\right)  = \sum_n 
u_{\bot,n}\left( t\right) \,e^{in\phi }$,   
$u_{\parallel }\left( \phi ,t\right)  = \sum_n 
u_{\parallel ,n}\left( t\right) \,e^{in\phi }\,,$  
and use the above described cutoff procedure of (\ref{cutoffprescr}) 
respectively (\ref{effective cut-off}). We then obtain the linearized 
equations of motion for parallel and perpendicular oscillations 
of the filament 
\begin{eqnarray} \partial _{t}u_{\parallel ,n} & = & b_{n}\,u_{\bot ,n}\,,
\nonumber\\
\label{eqmotionforu}
\partial _{t}u_{\bot ,n} & = & -a_{n}\,u_{\parallel ,n}\,.
\end{eqnarray}
The coefficients in this  
linearized version of (\ref{BS equation of motion}),  
\begin{eqnarray}\label{anbndef}
a_n & = &\frac{\Gamma }{4\pi }\frac{1}{%
r_{0}^{2}}\left[ n^{2}\ln \cot \frac{\delta }{4}\,-I_{\parallel ,n}\right] \,,
\\
b_n& = &\frac{\Gamma }{4\pi }\frac{1}{%
r_{0}^{2}}\left[ \frac{1+\cos(n\delta)}{2\cos \frac{\delta }{2}}
-\left( 1-n^{2}\right) \ln\cot 
\frac{\delta }{4}-I_{\perp ,n}\right],\nonumber
\end{eqnarray}
are given in terms of 
integrals containing, due to our parametrization (\ref{shape ansatz})
of the ring oscillations, 
trigonometric functions only:  
\begin{eqnarray}
I_{\perp ,n}
& = &\frac{1}{8}\int_{\delta }^{2\pi -\delta }\!\!d\phi
{\left| \sin \frac{\phi }{{\small 2}}\right| ^{-3}}\! \left[
\left( 1-\cos \left( n\phi \right) \right) \frac{1}{2}\,\left( 1+\cos \phi
\right)   \right.\nonumber\\
& - &\left. n\,\sin \left( n\phi \right) \sin \phi +n^{2}
\left( 1-\cos \phi \right)
\right]\nonumber\\
&=&-\frac{1}{2}\left[ \frac{n\sin \left( n\delta \right) }{\sin \left( \frac{%
\delta }{{\small 2}}\right) }-\,\frac{\sin ^{2}\left( n\frac{\delta }{%
{\small 2}}\right) \cos \left( \frac{\delta }{{\small 2}}\right) }{\sin
^{2}\left( \frac{\delta }{{\small 2}}\right) }\right] \nonumber\\
& &\qquad + \left( 2n^{2}-\frac{1%
}{2}\right) \sum_{j=1}^{n}\frac{\cos \left[ \left( j-\frac{1}{2}\right)
\delta \right] }{2j-1}\,,\nonumber\\
I_{\parallel ,n} 
& = & I_{\perp,n}-\frac{1}{8}\int_{2\pi -\delta }^{\pi }d\phi 
\frac{1-\cos \left( n\phi \right) }{%
\sin \frac{\phi }{{\small 2}}}\nonumber\\
&= & \label{Idef} 
I_{\perp,n}
- \sum_{j=1}^{n}\frac{\cos \left[ \left( j-\frac{1}{2}\right)
\delta \right] }{2j-1}\,.
\end{eqnarray}

In Fig. \ref{abcoeff}, the coefficients $a_n$ and $b_n$, for $r_0/\xi_c=7.5 $, 
are shown.  
\begin{center}
\begin{figure}[hbt]
\begin{center}
\begin{figure}[hbt]
\psfrag{an}{\large $a_n$}
\psfrag{bn}{\large $b_n$}
\psfrag{n}{\large $n$}
\psfrag{r0xic60}{\large $r_0/\xi_c=60$}
\psfrag{delta0}{$\delta\rightarrow 0$}
{\includegraphics[width=0.48\textwidth]{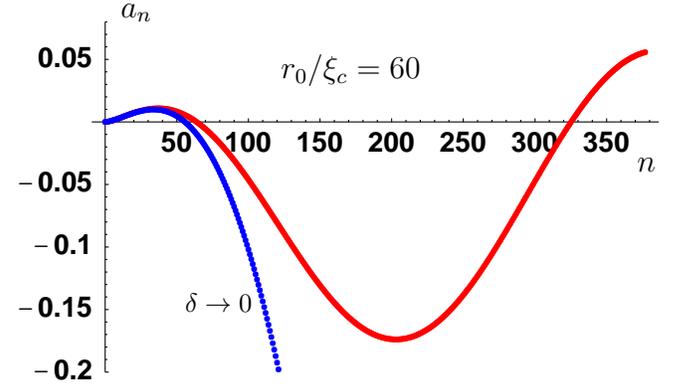}}
\vspace*{0.5em}
\caption{\label{anFig} The coefficients $a_n$ given in Eq. (\ref{anbndef}), 
 up to $n=2\pi\delta^{-1}\simeq 2\pi r_0/\xi_c$, for $r_0/\xi_c=60$, 
in units of $\omega_c$ defined in Eq. (\ref{cyclo}).
For these small values of $\delta$, $a_n$ and $b_n$ are 
indistinguishable within the Figure's resolution.  
The dotted curve is the asymptotic result for $\delta\rightarrow 0$ 
in equation (\ref{ablimes}).}
\end{figure}
\end{center}

\psfrag{anbn}{\large $(a_n,b_n)[\omega_c]$}
\psfrag{an}{\large $a_n$}
\psfrag{bn}{\large $b_n$}
\psfrag{n}{\large $n$}
\psfrag{r0xic}{\large $r_0/\xi_c=7.5$}
{\includegraphics[width=0.48\textwidth]{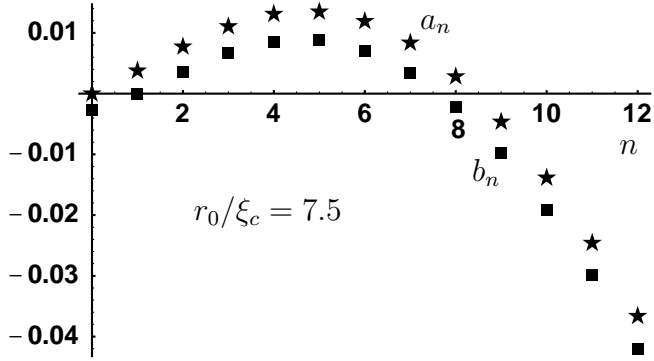}}
\vspace*{0.5em}
\caption{\label{abcoeff} The coefficients $a_n$ (stars) and $b_n$ (boxes)
in the equations of motion (\ref{eqmotionforu}), in units of
 $\omega_c$ defined in Eq. (\ref{cyclo}), for $r_0/\xi_c=7.5$, up to $n=12$.
 There occurs a dynamically unstable
 mode of imaginary frequency for $n=8$.}
\end{figure}
\end{center} 
According to (\ref{eqmotionforu}), the 
frequencies of oscillation of a vortex ring evaluate from 
\begin{equation}\label{dispersion}
\omega^2_{n} 
=a_{n}b_{n}\,.
\end{equation}
We scale the coefficients $a_n,b_n$ in Figs. \ref{abcoeff},\ref{anFig}  
and frequencies $\omega_n$ in Fig. \ref{ringmodes} below   
in units of the fundamental 
cyclotron frequency of the vortex core,  
\begin{equation}
\omega_c=\frac{4|\Gamma|}{\pi \xi_c^2}\,, \label{cyclo} 
\end{equation}
the frequency with which a point on the core 
revolves around the line designated by $\bm R(\phi,t)$ (the name stemming
in the magnetic analogy from the role of $\Gamma$ as a flux). 

The frequencies (\ref{dispersion}) 
are exactly zero for both $n=0$ and $n=1$, to any order in $\delta$. 
In the first case, $a_0=0$ and $b_0=(\Gamma/4\pi r_0^2)(\cos(\delta/2))^{-1}-
v_0/r_0$, 
whereas  in the latter case,
$a_{1}=v_0/r_0$ and $b_{1}=0$.
The first case of symmetry 
corresponds to $u_{\bot,0}=$const., and tells us that 
the radius of the ring as a function of $\delta$
is determined up to the (constant) value of $u_{\bot,0}$. 
In the second case, in turn, 
$u_{\parallel,1}$ is a constant. The resulting 
line deformation resembles, for $\delta \rightarrow 0$,
 a translation of the ring as a whole. 
Except for radii $r_0$ which are just about an order
of magnitude above $\xi_c$, the coefficients $a_n$ and $b_n$ 
are practically the same over the range of allowed 
values of $n$, so that for $\delta \ll 1$ the frequency  
$\omega_n$ is essentially equal to either $|a_n|$ or $|b_n|$, and the 
waves around the ring are circularly polarized, like for ordinary 
Kelvin waves. However, for $r_0$ getting
closer to $\xi_c$, $a_n$ becomes increasingly different from $b_n$, and the 
waves become elliptically polarized, the absolute 
ratio of amplitudes in the ring plane 
and out of the plane being given by $|u_{\bot ,n}/u_{\parallel,n}| 
= \sqrt{|a_n/b_n|}$. A small ring thus oscillates 
more  in the ring plane than out of the ring plane.    
Let us also stress that for the calculation of the oscillation modes,    
the nonlocality in the Biot-Savart
integrals of (\ref{BS energy}) respectively (\ref{BS equation of motion}),  
is fully taken into account.    
For direct comparison with the 
dispersion of Kelvin waves on rings, we
consider, for fixed $n$, 
the limes 
$\delta \rightarrow 0$ in (\ref{anbndef}) and (\ref{Idef}),  to obtain 
\begin{eqnarray} 
a_n &=&
\frac{\Gamma }{4\pi }\frac{1}{%
r_{0}^{2}}\left[n^2\left(\ln \left[\frac{4r_0}{\xi_c}\right]
-2 S_n +\frac12\right)
+\frac32 S_n\right]\,,
\nonumber\\
b_n & = & \frac{\Gamma }{4\pi }\frac{1}{%
r_{0}^{2}}\left[
\left(n^2-1\right)\left(\ln \left[\frac{4r_0}{\xi_c}\right]
-2 S_n +\frac12\right)\right.\,,\nonumber\\
& &\left. \qquad\qquad -\frac32 \left(S_n-1\right)\right]
\qquad (n \ll 1/2\delta)
\label{ablimes}
\end{eqnarray}
where $S_n = \sum_{j=1}^{n}
({2j-1})^{-1}$. 
There is an important difference of the  dispersion relation (\ref{dispersion})
(which is exact within our hollow core model), 
with $a_n$ and $b_n$ from (\ref{anbndef}),    
to the usually quoted asymptotic results 
of Lord Kelvin, J. J. Thomson \cite{Kelvin,JJThomson}
 and Grant \cite{grantModes,extract} (also cf. the work 
of Pismen and Nepomnyashchy \cite{pismennepo}, which found the 
same result as Grant, but within a much simpler scheme, similar to ours). 
These last results
correspond to the relations (\ref{ablimes}) for the coefficients 
$a_n$ and $b_n$  in the limit 
of $\delta \rightarrow 0$ for fixed $n$ 
(save for different core structure constants).  
The important difference consists in the fact that the geometric cutoff 
prescription (\ref{cutoffprescr}), which ensures that 
a core of diameter $\xi_c$ 
is always excluded in the evaluation of 
(\ref{BS equation of motion}), is taken care of in relations 
(\ref{anbndef}) exactly for any admissible value of the ratio 
of ring radius and core diameter $r_0/\xi_c$,  
such that $n\delta\sim O(1)$ can be
consistently realized.  
The anomalous branch also occurs, shifted to smaller mode numbers, 
as a consequence of relations (\ref{ablimes}).  
However, the minimum resides at values of $n\sim \delta^{-1}$ 
which are beyond the applicability of equations (\ref{ablimes}).
We have depicted the difference between the exact and asymptotic 
results in Fig. \ref{anFig}, for  
the whole range of $n$ up to the value 
$n=2\pi\delta^{-1}\simeq 2\pi r_0/\xi_c$, corresponding 
to $k\simeq 2\pi /\xi_c$ ($\lambda\simeq \xi_c$). For the value 
$r_0/\xi_c=60$ used, the coefficient 
$b_n$ is essentially identical to $a_n$
within the resolution of the Figure (cf. Fig. \ref{abcoeff}, which has
$r_0/\xi_c=7.5$, and where the difference between 
$a_n$ and $b_n$ is clearly discernible).  

For completeness, we state the solution of the equations of motion 
(\ref{eqmotionforu}). 
If the vortex ring undergoes at time $t=0$ a deformation represented by 
\begin{eqnarray}
u_{\parallel }\left( \phi ,0 \right) 
& = &\Re \sum_n u^0_{\parallel ,n}  
e^{in\phi } 
\nonumber\\ 
u_{\perp }\left( \phi ,0\right)
& = &\Re \sum_n u^0_{\bot ,n} 
e^{in\phi } \,, 
\end{eqnarray}
the solution at a later time $t$ takes the form 
\begin{eqnarray}
u_{\parallel }\left( \phi ,t\right) & = & \Re \sum_n 
\left[ 
u^0_{\parallel ,n}
\cos \left( \omega_n t\right)
+u^0_{\bot ,n} 
\,\frac{b_{n}}{\omega_n}\,
\sin \left(  \omega_n 
t\right) 
\right] e^{in\phi }
\nonumber
\\ 
u_{\perp }\left( \phi ,t\right)
& = & \Re \sum_n\left[
u^0_{\bot ,n} 
\cos \left(  \omega_n t\right)
-u^0_{\parallel ,n} 
\,\frac{a_{n}}{\omega_n}\,\sin \left(  \omega_n
t\right) 
\right] e^{in\phi }
\,,\nonumber\\ 
\end{eqnarray}
where $\omega_n=\sqrt{a_n b_n}$. 
\section{Hamiltonian}  
\label{Hamiltonian} 
The Hamiltonian corresponding to the canonical variables 
(\ref{canonical variables}) and the action (\ref{vortex action}) 
assumes the form 
\begin{eqnarray}
{\cal H} & = & E_0 
\label{totalenergy}
+ \sum_{2\le n\le n_c}
\frac{\rho_0 \Gamma r_0}2\left[ a_n u_{\parallel,n}^2+
b_n  u^2_{\bot,n}\right]\,,
\end{eqnarray}
where the stationary energy of the ring is given by 
\begin{equation}
E_0 =  \frac{\rho_0 \Gamma^2 r_0}{2}\left( \ln \left[
\cot\left(\frac\delta 4\right)\right]-2\cos\left[\frac\delta 2\right]
\right)
\,. \label{E0}
\end{equation}
The (constant) variables $u_{\bot,0}$ and $u_{\parallel,1}$ 
do not appear in (\ref{totalenergy}) 
because of our choice of parametrization (\ref{shape ansatz}), which 
corresponds 
to a transformation to the rest frame of a ring of radius $r_0$, moving 
with velocity $v_0$. The above expression for $\cal H$, then, 
represents the rest frame Hamiltonian of the vortex.
The phase space variables may, for example, be chosen to be   
$q_n= u_{\parallel,n}$ and $p_n = \rho_0\Gamma r_0 u_{\bot,n}$, so that the
mass $M_n= \rho_0\Gamma r_0 /b_n$ and elasticity (spring) constant  
$D_n = \rho_0\Gamma r_0 a_n$. Equally well, we may choose 
the 
option 
$q_n= u_{\bot,n}$ and $p_n = -\rho_0\Gamma r_0 u_{\parallel,n}$, 
which reverses the role of mass and elasticity coefficients 
in conventional Hamiltonian language (replaces $a_n$ by $b_n$ and 
$b_n$ by $a_n$ in $M_n$ and $D_n$). The identity of phase space and 
configuration space \cite{poincare,onsager,Regge,annalspaper}
implies that both options are viable. 
\begin{center}
\begin{figure}[hbt]
\psfrag{w}{\large $\omega_n$}
\psfrag{n}{\large $n$}
{\includegraphics[width=0.48\textwidth]{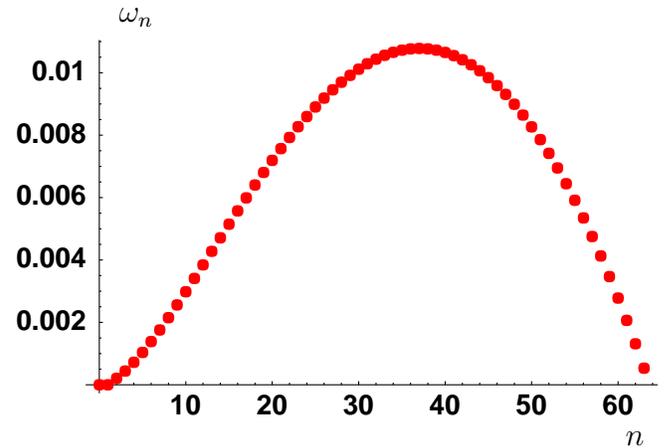}}
\vspace*{0.5em}
\caption{\label{ringmodes} Stable 
oscillation frequencies of a vortex ring,  
in units of the cyclotron frequency 
$\omega_c$, 
as a function of the mode number $n$, 
for the ratio $r_0/\xi_c=60\simeq \delta^{-1}$, up to $n_c\simeq \pi /
(3 \delta)$.
For this ratio of radius and core diameter, 
the exact value is $n_c=63$. 
The maximum is, essentially independent from the value of $\delta$, 
situated at max[$\omega_n] \simeq 0.011
\,\omega_c$, with $n\simeq n_c/2$. 
}
\end{figure}
\end{center}
From the Hamiltonian (\ref{totalenergy}), we gather 
that stable oscillation modes are those which have 
$a_n$ and $b_n$ {\em both} positive. 
They contribute positive energy to the Hamiltonian. 
Energetically unstable, though giving a real frequency,  
are modes which have $a_n$ and $b_n$ both negative, 
because they contribute negative energy to the Hamiltonian.
A different sign of $a_n$ and $b_n$ leads to dynamically unstable modes,   
which have imaginary frequencies and amplitudes ${\bm u}$ 
exponentially growing (or decaying) in time. 
For mode numbers above 
\begin{equation}
n=n_c\simeq \frac{\pi}{3\delta} \,,
\end{equation} 
and up to $n\simeq 5\delta^{-1}$, both $a_n$ and $b_n$ become 
negative, such that the energy contribution corresponding to these 
modes is {\em negative}. To quadratic order in the oscillation amplitude, 
oscillations with wavelengths smaller than 
 $\lambda \sim 6\xi_c $ thus imply 
that vortex modes of such small wavenumber are unstable. Hence, 
the stable spectrum is restricted to mode numbers of magnitude less than 
$n_c \simeq \delta^{-1}$, by definition the last mode number for which the
oscillation energy is positive semi-definite, before entering 
the negative energy domain seen in Fig. \ref{anFig}.
We have plotted the dispersion relation of the stable modes 
in Fig. \ref{ringmodes}, for $r_0/\xi_c=60$.  


With regard to the validity of the assumption of an incompressible fluid, we
note that the frequency at the maximum in Fig. \ref{ringmodes}, situated 
at $n
\simeq n_c/2\simeq (2\delta)^{-1}$ 
for all values of $r_0/\xi_c$ not too close to unity, 
scales as $\omega_n\simeq 0.011 
\,\omega_c$\,, 
with the cyclotron frequency $\omega_c$ defined in (\ref{cyclo}) 
($\omega_c\sim 10^{12}$\,sec$^{-1}$ in superfluid $^4\!$He (helium II)). 
Oscillation 
velocities thus remain, for moderate oscillation amplitudes of the order
of a few $\xi_c$, well below the speed of 
sound even in the superfluid helium II, where $\xi_c$ is of atomic 
size and $c_s\xi_c\sim \Gamma
=\pi\omega_c(\xi_c/2)^2 $, so that the 
incompressibility approximation holds.
This is more questionable for the second, much larger, frequency maximum at 
$n\simeq 3 \delta^{-1}$, corresponding to the maximum negative 
value of the coefficient $a_n$ in Fig. \ref{anFig}, 
given by $a_n \simeq -0.18 \,\omega_c$.  

\vspace*{-1.5em}  
\begin{center}
\begin{figure}[hbt]
{\includegraphics[width=0.4\textwidth]{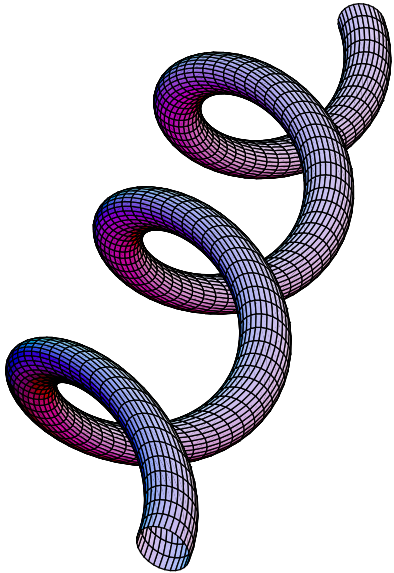}}
\vspace*{-0.5em}
\caption{\label{CoreCutoff} 
Shape of helical wave traveling along the ring, 
for mode numbers 
near the critical mode number $n_c$, showing  
the vortex core displacement and the volume exclusion 
caused by this displacement.  
}
\end{figure}
\end{center}
Consider, for a physical interpretation of the energetical 
instability around $n=n_c$, 
Fig. \ref{CoreCutoff}, where we show the shape of deformation  
of the vortex core for a small wavelength of order 
$\lambda_c\simeq 6\xi_c$, 
corresponding to the crossover to the unstable 
oscillations r\'egime.  We may infer that the
negative oscillation energy, occuring at a smaller wavelength than that  
shown in Fig. \ref{CoreCutoff}, 
is due to a volume exclusion effect. The excluded
core volume kinetic energy is large enough such that there is only small  
energetical cost of exciting a perturbation on the filament for the stable
modes with mode numbers slightly below $n_c$, 
 and an energetical gain for the unstable ones.
In the Hamiltonian (\ref{totalenergy}),
we neglect the tail of positive excitation energies 
corresponding to positive values of $a_n,b_n$ seen in Fig. \ref{anFig}, 
because it is very close to the limit of core elements 
touching themselves, at which point 
our core model certainly becomes invalid, 
because it is then meaningless to speak about oscillations of a hollow 
torus. 
That the coefficients $a_n$ and $b_n$,  
and thus the excitation energy, do increase again after 
$n\simeq 3 \delta^{-1}$, can be traced back to the fact that line 
elements having like circulation approach each other closely 
if we further 
compress the spiral of Fig. \ref{CoreCutoff} along its axis.  
The volume energy exclusion 
effect we just described 
is then counterbalanced, for these very short wavelengths,  
by the resulting strong repulsion of adjacent 
elements of the same circulation. 

The frequency (\ref{dispersion}) is imaginary, if $a_n$ and $b_n$ have 
different signs, and a {\em dynamical} instability results 
\cite{widnallsullivan73,pismennepo}. 
We stress, however, 
that the unboundedness of the Hamiltonian (\ref{totalenergy}) 
from below leads to the {energetical} instability of the ring for 
mode numbers beyond $n_c$. This instability will exist for any value of 
$\delta$ respectively of $r_0/\xi_c$. 
The change induced by choosing some different, more 
regular and differentiable 
core structure than our prescription (\ref{effective cut-off}) 
is the numerical value of the  slope of the negative group velocity
branch, within a number of order unity, and the point, 
as a function of $r_0/\xi_c$, where $n=n_c$ and the 
excitation energy becomes negative. The fact 
that around $n_c$ there are dynamical instabilities taking place has
also been recognized in \cite{pismennepo}, 
where the dynamical instability was investigated using the 
Gross-Pitaevski\v\i\, model of a superfluid, and it was indeed found that 
$n_c\xi_c/r_0 = O(1)$. 
However, what has been missed in this work (and others in 
the conventional fluid mechanical framework 
\cite{saffman,widnallsullivan73}), is that 
an energetical instability will take place right after we have crossed the 
dynamical instability region, independent 
from the precise value of the critical
$n_c$ as a function of $r_0/\xi_c$.   
It is apparent from Fig. \ref{CoreCutoff} that the instability 
will persist
for any (regularizing) model taken of the core region, as long as the energy
density of the core is significantly less than that of the surrounding
bulk fluid, {\it i.e.}, as long as it is still significantly 
reduced compared to the bulk if we, for example,  
turn on interactions (take into account compressibility) 
inside the core. For smaller energy density differences between 
core and bulk, the value of $n_c$ will be shifted upwards 
(for a given value of $r_0/\xi_c$), but the energetical instability 
will still exist. 
\section{Low temperature 
thermodynamics of vortex oscillations in helium II}
\label{Thermodynamics} 
Up to this point, our considerations have been in terms of a classical vortex. 
Consider now the quantum mechanical zero point fluctuations of a vortex line 
in the quantum fluid helium II, 
each with a contribution $\frac12\hbar \omega_n$ 
to the vortex (ground state) energy. 
If we sum up these contributions to the limiting value $n_c$, we get
$E_{\,\rm fl}\equiv \sum_{n=2}^{n_c}\frac12 \hbar\omega_n
\simeq 0.0035\,\hbar\omega_c\, \delta^{-1} $
(cf. the area under the dispersion curve in Fig. \ref{ringmodes}).
Comparing this with the stationary energy $E_0$, see equation (\ref{E0})
of an undeformed ring in helium II,  
we obtain 
\begin{equation}
{E_{\,\rm fl}}
\simeq 0.014
\left(\frac{d}{\pi^{2/3}\xi_c}\right)^3 
\frac{E_0}{\ln (r_0/\xi_c)}\,,  
\end{equation}
where the interparticle distance $d=(\rho_0/m)^{-1/3}$ 
($\sim \xi_c$ in helium II).  
The total quantum mechanical fluctuation energy in the stable modes
(at zero temperature)  
is thus much less than the stationary energy $E_0$, of the order of percent. 
This need not be the case if we take into account 
thermal fluctuations as well.
The 
vortex free energy
may be written 
\begin{eqnarray}
{\cal F}(T, \delta) & = & E_0+ 
\beta^{-1}\sum_{n=2}^{n_c} 
\ln \left(2\sinh[\beta \hbar\omega_n/2]
\right)
\,. 
\end{eqnarray}
The entropic part of the free energy, due to ring oscillations,  
plays an important role if the temperature is a significant fraction of 
the cyclotron energy of the core.
We stress that 
the temperatures for fluctuations to 
become important are significantly higher if the cutoff is 
chosen well below $n_c$. 
We have also convinced ourselves that the absolute 
ratio of the oscillation free energy part over the stationary ring energy,  
$|{\cal F}-E_0|/E_0$, is the larger the smaller the radius $r_0$ is, 
{\it i.e.}, the oscillations play an increasingly important thermodynamic
r\^ole for smaller rings. 

For low temperatures, fulfilling 
$k_B T \ll  {\rm max}\,[\hbar\omega_n] \simeq 0.011\, \hbar\omega_c$\,, 
Kelvin modes of small wavenumber  
{and} the modes with approximately linear dispersion 
around $n_c$ (cf. Fig. \ref{ringmodes}), are populated 
\cite{BDVPhysFluids85}.   
In the superfluid helium II, where $\hbar\omega_c$ is of the 
order ten Kelvin, 
the condition on the temperature leads to the requirement  
$T \ll 100$ mK, which is feasible in experimental practice.
For vortex rings of a given size and orientation, we thus expect two 
contributions to the specific heat at low temperatures, coming
from the aforementioned two asymptotic 
branches of excitations on the filament.  
For the indicated range of mode numbers, we 
may approximate the dispersion by the Kelvin-like form 
\begin{eqnarray}
\omega_K & = &
\gamma_1  n^2 \ln [{n_c^*}/{n}]\qquad \left(1<n \ll n_c/2\right)\,,
\end{eqnarray}
where the parameters $\gamma_1$ 
and $n_c^*$ 
are 
\begin{equation}  
\gamma_1 
= \frac{|\Gamma| }{4\pi r_0^2}
= \omega_c \left(\frac{\xi_c}{4r_0}\right)^2\,, \quad\qquad 
n_c^*= 8\sqrt e\, \frac{r_0}{\xi_c}\,. 
\end{equation} 
Near $n_c$, a linear law obtains:  
\begin{equation}
\omega_A \simeq \gamma_2 (n_c-n) \qquad 
\left(n_c-n \ll n_c/2\right) \,,
\end{equation}
where, numerically,
\begin{equation}
\gamma_2\simeq 0.045\, \omega_c \, {\xi_c}/{r_0}\,.
\end{equation} 
Both of these approximate 
dispersion relations are valid for large $r_0/\xi_c$ (small $\delta$).
The 
density of states for the $\omega_K$-branch, within 
logarithmic accuracy, may be written   
$N_K(E)\simeq (4\hbar \gamma_1 \ln[n_c^*]\, E)^{-1/2}$; 
for the $\omega_A$-branch it is independent of the energy $E$,  
$N_A (E) = (\hbar\gamma_2)^{-1}$.  
The asymptotical behaviour of the vortex specific heat for low temperatures 
then assumes the form 
\begin{eqnarray}
\frac{C_{v}}{k_B} 
& \simeq &%
\frac{2\pi r_0}{\xi_c}
\left( 
\frac{1.1}{\sqrt{\ln n^*_{c}}}
\,\sqrt{\frac{k_B T}{\hbar\omega_c}}
+ 11.6\,  \frac{k_B T}{\hbar\omega_c}
\right)
\,.
\end{eqnarray}
It is, as expected, proportional to the ``volume'', {\it i.e.} 
the circumference of the 
ring, and  
has a contribution proportional to $\sqrt T$ from the Kelvin-like modes and 
a new contribution proportional to $T$ stemming from the linear dispersion,  
large wavenumber branch. 
This last term 
gives a dependence of $C_{\rm L}$ proportional to the area 
$2\pi r_0\xi_c$.  
For a randomized ensemble of vortex rings, with different orientations 
and radii, we expect the indicated dependence on temperature to hold 
for a dilute system of effectively noninteracting rings. For a dense vortex 
tangle, coupling of the vortex rings by mutual induction   
will modify the spectrum and the above thermodynamic behaviour, 
a problem 
which is left for future work.     

\section{Conclusion} 
We have derived 
the oscillation modes on vortex rings, 
using the canonical phase space structure of small ring 
oscillations in an incompressible, inviscid fluid and a geometrical 
cutoff procedure for the core region. 
Beyond a critical wavenumber, the excitation energy becomes negative, 
indicating that the vortex ring is energetically unstable for perturbations 
on scales of short wavelengths. The instability relies on the energy 
exclusion effect of the deformed core, and may be interpreted in conventional 
Hamiltonian language as being due to the fact that the classical or 
quantum particle representing the excitation has both  
a kinetic energy with negative mass and a potential with 
negative spring constant. 
 

The existence and peculiar anomalous dispersion 
of propagating modes with very small wavelengths 
should have important implications for the dynamics 
of vortex reconnection events \cite{svistunov},  
as well as the final stages of the energy cascade process 
in superfluid turbulence \cite{vinen2000}. In addition, 
we expect that scattering cross sections of the elementary 
roton excitation in helium II with vortices, 
and thus the coefficients of mutual friction between superfluid and normal 
components \cite{vinen57}, will be influenced by the presence of low energy modes 
with wavenumbers of order the inverse core size.   
\begin{acknowledgments}
The work of 
U. R. Fischer 
was financially supported by the DFG (FI 690/1-1).
\end{acknowledgments}


\begin{references}
\bibitem{extract} 
From the calculations in \cite{grantModes}, 
the relation of the core diameter $\xi_c$ in our equation 
(\ref{cutoffprescr}) to the 
Gross-Pitaevski\v\i\, coherence length $\xi_{\rm GP}$, using 
the expression (\ref{v0}) for the propagation velocity, 
is given by  
$\xi_{\rm GP}/\xi_c = 2/\exp[0.615] \simeq 1.08$.
\end{references}
\end{document}